# Electrode Geometry Optimization in Vortex-Type Seawater Magnetohydrodynamic Generators

Arleen Natalie[1], Budiarso[1], and Ridho Irwansyah[1]

[1] Department of Mechanical Engineering, Faculty of Engineering, Universitas Indonesia, Indonesia

## Abstract

Magnetohydrodynamics (MHD) generators present a promising pathway for clean energy conversion by directly transforming conductive fluids' kinetic energy into electricity. This study investigates the impact of electrode geometry modifications on the vortex-type seawater MHD generator's performance. Three electrode designs: partial, whole-area, and spiral; were analyzed through combined analytical and numerical simulations using COMSOL Multiphysics. This study focuses on internal resistance reduction, current density distribution, and overall power output. The results indicate that the electrode area and spacing are critical determinants of performance. The whole-area electrode achieved the highest output, with a 155% increase in power compared to the baseline partial electrode. The spiral electrode demonstrated reduced internal resistance and improved current flow but exhibited lower open-circuit voltage due to reduced electrode spacing. The simulations showed strong agreement with the theoretical models, with a deviation of less than 4% in the open-circuit voltage predictions. These findings highlight the importance of geometric optimization for advancing seawater-based MHD generators as sustainable and efficient energy conversion systems. By integrating vortex chamber design, seawater as a working fluid, and optimized electrode geometry, this research contributes to the development of next-generation MHD technologies that address efficiency and scalability challenges for clean energy applications.

**Keywords**: magnetohydrodynamics, vortex generator, seawater, electrode geometry, clean energy, hydropower.

# I     Introduction

The global energy demand is projected to rise significantly as humanity advances toward a highly technological and sustainable civilization. Meeting these demands requires innovative energy conversion technologies that minimize environmental impact while maximizing efficiency. In the presence of a magnetic field, magnetohydrodynamics (MHD) generators convert the kinetic energy of conductive fluids, such as plasma or seawater, into electrical energy. Unlike conventional generators, MHD devices eliminate mechanical moving parts, potentially reducing maintenance and increasing durability. Despite decades of research, MHD generators face challenges that reduce efficiency, including flow instabilities, internal resistance, and end-effects. Recent studies have explored vortex chamber configurations to improve flow stability and novel electrode arrangements to mitigate resistance and enhance current collection. Seawater is an abundant and non-toxic conductive medium and a practical working fluid for MHD applications. This study investigates how electrode geometry influences the performance of a vortex-type seawater MHD generator, focusing on efficiency improvements and design trade-offs.

Previous research on MHD generators has explored various chamber designs and their influences on fluid conductivity. Cylindrical chambers provide smooth and symmetrical flow paths, but they often suffer from flow separation and uneven magnetic field distribution, thereby reducing efficiency. For example, Li [1] reported that while partial disk channels enhance plasma uniformity and reduce energy losses, efficiency declines sharply at angles above 15°. Fujino *et al.* [2] showed that asymmetric one-sided inflow ducts can match symmetric designs in energy extraction but introduce asymmetric plasma distributions requiring additional length for stabilization. In annular generators, Pérez-Orozco and Ávalos-Zúñiga [3] highlighted that the internal resistance depends on both the Hartmann number and duct radius, which must be matched to the load resistance. Bühler *et al.* [4] observed that liquid-metal flows in ducts with Faraday current insulation suffer from localized current densities and pressure drops at wall boundaries. Fonseca *et al.* [5] demonstrated through 3D simulations that circular ducts with Hartmann numbers >10 can reduce flow velocity by up to 35%, although the effects are less pronounced under simpler field configurations. Kobayashi *et al.* [6] showed that a nonuniform magnetic flux density produces an M-shaped velocity profile, whereas Hu *et al.* [7] linked this to thin Hartmann layers and end-effects. Liu *et al.* [8] confirmed similar leakage and eddy current issues in wave-energy MHD setups. Cosoroabal *et al.* [9] noted that magnet misalignment in rectangular ducts caused off-centered streams and stagnant fluid pockets, and Domínguez-Lozoya et al. [10] highlighted the significant end-effects and energy losses in the reciprocating channels. Blishchik et al. [11] observed flattened velocity profiles and boundary layer effects across MHD systems. More recently, Yang et al. [12] demonstrated that triangular strips in duct cross-sections can mitigate pressure drops, whereas Jiang et al. [13] showed that optimizing the external magnetic fields can reduce the end currents and Joule heating, thereby improving the generator efficiency by 9.5%.

The end-effect remains a major limitation in rectangular MHD generators, arising from nonuniform magnetic fields, finite magnet lengths, and geometric misalignments that promote energy losses through eddy currents and leakage currents near the electrodes. Although strategies such as external magnetic field optimization and the introduction of triangular strips have shown partial improvements, current duct designs remain inadequate due to persistent issues, including velocity disturbances, stagnant regions, and pressure drops.

The use of vortex chambers in MHD generators is a promising alternative. Unlike conventional rectangular ducts, vortex configurations minimize dead zones and promote balanced velocity distributions, resulting in more stable fluid–field interactions and improved energy conversion efficiency. Panchadar et al. [14] demonstrated that vortex MHD generators can achieve power densities of 34 W/cm³, with optimization to 102 W/cm³. Gupta et al. [15] demonstrated that Galinstan-based vortex systems can generate up to 3 W, and Gupta et al. [16] reported outputs up to 7 W with higher fluid conductivity. West et al. [17] noted that the internal resistance remains a critical barrier in compact and versatile vortex generators. Similarly, Ávalos-Zúñiga and Rivero [18] found that under turbulence, the flow distribution in DC vortex chambers can shift from parabolic to S-shaped profiles, influencing the overall performance. Collectively, these studies show that vortex geometries can mitigate end-effects and energy losses, offering a promising pathway for MHD technology advancement.

Liquid metals provide clear advantages for MHD generation, including high electrical conductivity and reduced end losses, but they also have significant limitations. Galinstan, for example, is effective yet costly and limited in availability, whereas traditional metals such as mercury and lead, although previously employed in studies such as Ryan et al. [19] on a vertical gravity-driven MHD generator achieving over 0.5 MW/m³, are toxic and environmentally hazardous. Plasma has also been investigated as a working fluid because of its excellent conductivity; however, it requires extremely high temperatures and complex containment. Tang et

*al.* [20] showed that while Faraday-type plasma MHD generators can operate efficiently, their performance is constrained by the high thermal input required, which reduces their practicality. In contrast, seawater is a viable alternative. Although its conductivity is lower than that of metals or plasma, it is abundant, nontoxic, and inexpensive. Seawater covers 71% of the Earth's surface and comprises approximately 1.3 billion cubic kilometers of volume. It is a readily available resource that can overcome the limitations of liquid metals and plasma, making it a compelling candidate for MHD energy harvesting.

A multifaceted approach is required to address MHD generator challenges. The implementation of vortex chamber designs can effectively mitigate end-effects and enhance flow stability, thereby optimizing the energy conversion process. Using seawater as the working fluid presents a practical advantage compared with other fluids, such as liquid metals or plasma, due to its abundance, nontoxicity, and cost-effectiveness. Furthermore, refining electrode geometry to reduce internal resistance and improve current distribution is crucial for boosting the overall efficiency and performance. This study uniquely integrates three innovative strategies: vortex chamber optimization, seawater utilization, and improved electrode design; highlighting a novel approach that has not been explored in previous research. By addressing these interconnected aspects, this research aims to bridge existing gaps and significantly advance MHD technology, driving progress toward more efficient and sustainable power generation solutions.

## II  Method

### II.1  Analytical method

Three main factors influence electromagnetic induction: fluid motion (u), magnetic field (B), and resulting current (J). According to Fleming's right-hand rule, these interactions are illustrated in Fig. 1. The interaction between the fluid motion (u) and the magnetic field (B), which generates an induced electric field, is the key phenomenon in electromagnetic induction. This field integrates along a line to produce the electric potential or voltage (V). The experimentally measured open circuit voltage $V_{oc}$, measured experimentally, is compared with the theoretical voltage derived from formulas. Measured and theoretical voltage values are analyzed and compared. The current density (J) is calculated from the electric field $E_{ind}$ and electrical conductivity (σ), considering the area and current I. The analysis concludes with the calculation of the electrical power $P_e$ generated by the MHD generator.

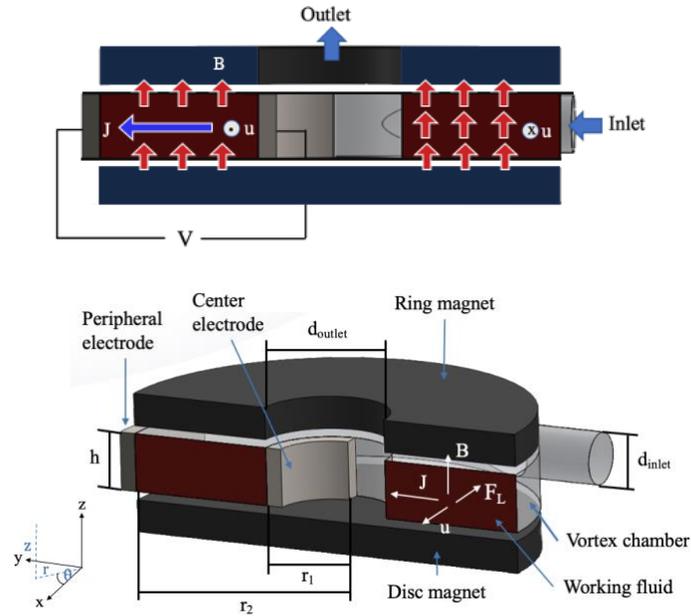

Figure 1. Working Fields and Orientation on Vortex MHD Generator

The study evaluates both internal and external resistances, incorporating them into the calculations for current and power. The relationship between the parameters is depicted in Fig. 2. The model will be simulated using COMSOL Multiphysics.

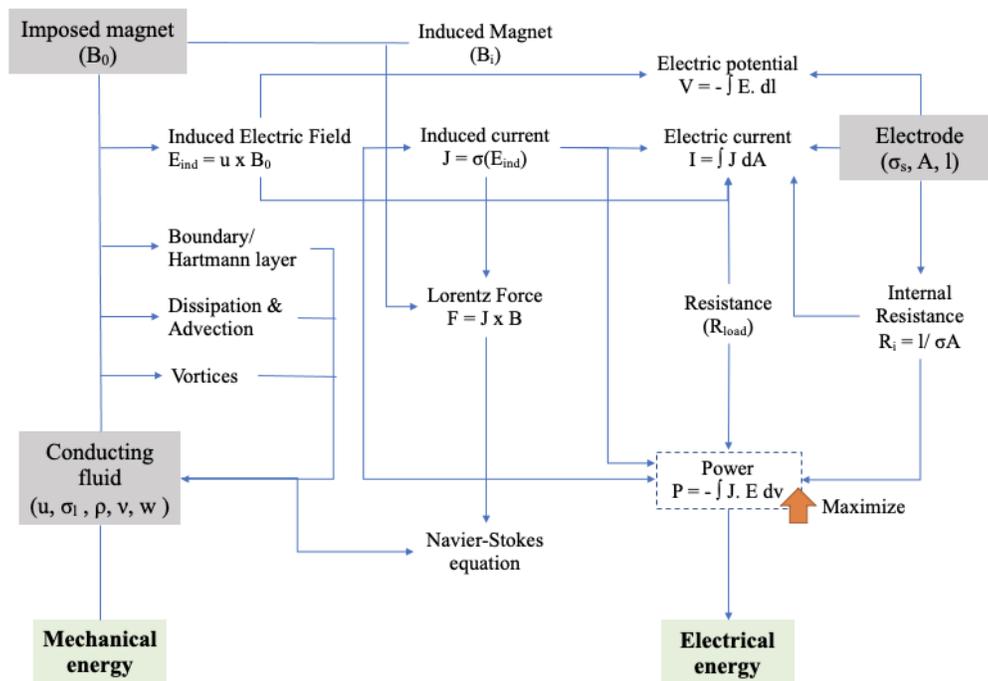

Figure 2. Parametric Relation in Magnetohydrodynamics Generator

In the MHD generator, the fluid enters through the inlet, circulates through the channel, and exits at the outlet. The cylindrical coordinate system accommodates the tangential motion of the fluid in the theta ($\theta$) direction, while the magnetic field is applied in the axial (z) direction, and the generated current flows radially (r). The interaction of the components will be analyzed analytically using Ohm's Law and Faraday's Law, with fluid behavior examined through the continuity equation and Navier-Stokes equation. The analysis concludes with MHD, incorporating the Lorentz force effects.

## II.2   Optimization of electrode geometry

The geometric characteristics of the electrodes significantly impact the performance of the generator, particularly through parameters such as electrode spacing (l) and electrode area (A). Reducing l and increasing A can lower internal resistance ($R_i$), as shown in Equation (9). An increase in A also enhances the generated current (I) based on Equation 4.6, though there is a trade-off as reducing l can diminish the open-circuit voltage ($V_{oc}$) according to Equation (4). Several electrode geometry enhancements were simulated to assess their impact on power output. The designs include: the partial electrode with total area 0.0014 m², whole area electrode by expanding the peripheral electrode to cover the entire chamber, the area increased to approximately 0.0082 m², a 4.85-fold enhancement to partial electrode, and the spiral electrode, which aims to reduce the electrode spacing (l) while increasing A. The center electrode was modified into a spiral path with a 2.1-cm spacing and one full revolution. These design variations were analyzed to determine their effects on the electrical power output.

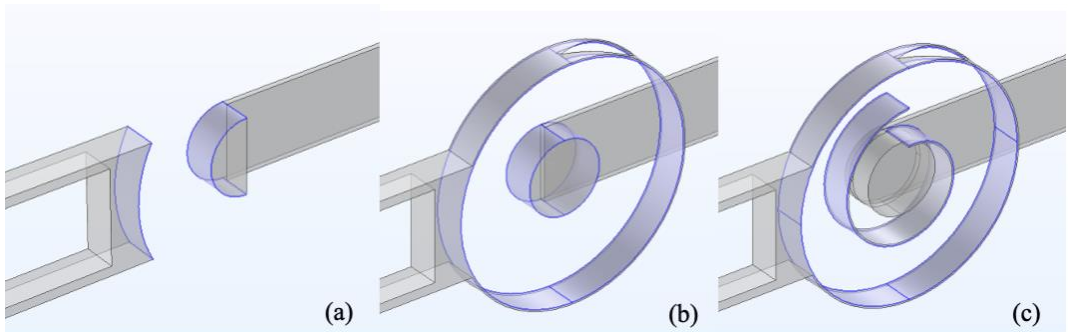

Figure 3. Electrode Geometry and Surface Current (blue) of (a) *Partial* (b) *Whole* and (c) *Spiral Electrode*

# III    Numerical Model

The COMSOL 3D model comprises four key components: free space, fluid, magnets, and electrodes. The free space represents the air surrounding the magnetic field and is large enough to encompass the magnets. The fluid, saltwater, is characterized by parameters such as electrical conductivity 5 S/m and inlet velocity ("$u_{inlet}$"), with its inlet adjacent to the free-space boundary. The magnets, consisting of a disc and ring, are positioned at the fluid's front and back, with their flux density applied according to orientation. The electrodes include a center electrode and a peripheral electrode, with the center electrode centrally located and the peripheral electrode along the fluid's side curvature. The boundary conditions for the electrodes are terminal and ground for one side, and electric insulation for the other. The load factor is set to its optimum level, 0.5 [21]. The material properties are detailed in Table 1.

Table 1. Material properties

| | *Free Space* | |
|---|---|---|
| 1. | *Relative permittivity* | 1 |
| 2. | *Relative permeability* | 1 |
| 3. | *Electrical conductivity [S/m]* | 0.01 |
| | *Fluid* | |
| 1. | *Relative permeability* | 1 |
| 2. | *Electrical conductivity [S/m]* | 5 S/m |
| 3. | *Relative permittivity* | 80 |
| 4. | *Density [kg/m³]* | 1025 |
| 5. | *Dynamic viscosity [Pa.s]* | $1.09 \times 10^{-9}$ |
| | *Magnet* | |
| 1. | *Relative permeability* | 1.05 |
| 2. | *Electrical conductivity* | 0 |
| 3. | *Relative permittivity* | 1 |
| | *Electrode* | |
| 1. | *Relative permittivity* | 1 |
| 2. | *Relative permeability* | 1 |
| 3. | *Electrical conductivity [S/m]* | $1.35 \times 10^4$ |

In the MHD simulation, three physics modules are used: Magnetic and Electric Fields, Turbulent Flow k-ω, and electrical circuit. The MHD multiphysics mode arises from the interaction between Magnetic and Electric Fields and Turbulent Flow k-ω, involving Lorentz and electromotive forces. The magnetic and electric fields are applied across the entire domain with boundary conditions, including Ampere's law, current conservation, magnetic insulation (defining boundaries for ground, terminal, and electric insulation), and surface current on both electrode surfaces. Ampere's law specifies varying remnant flux density between magnets.

The turbulent flow is applied to the fluid domain using the Reynolds Averaged Navier-Stokes (RANS) k-ω model, with boundary conditions set as the wall for the fluid boundaries, the inlet with parameter "$u_{inlet}$", and outlet with a static pressure of 0 Pa. The electrical circuit module measures the voltage generated by the MHD generator using the external coupling I vs. U. Simulations are conducted in both open circuit and loaded conditions. The loaded condition includes a resistor for external resistance, an ammeter for current measurement, and a voltmeter for voltage measurement.

A Physics-controlled mesh with sequence type is used in the simulation, using the default mesh sizes provided by COMSOL. The mesh includes element sizes such as coarser, coarse, normal, and fine, employed to conduct a mesh independence study by evaluating the fluid velocity deviations at the same points.

The study employed is a stationary analysis, where the flowing fluid is assumed to have reached a stable condition. A parametric study is configured by varying inlet velocity and magnetic field strength.

### III.1 Electromagnetic Analysis

The analytical derivation of the equations involves simplifying the velocity vector and magnetic field to uniform and scalar variables. According to Ohm's Law, the equations can be derived to obtain the electric field induction $\vec{E}_{ind}$ and $\vec{E}$ with equation

$$\vec{E}_{ind} = \vec{u} \times \vec{B} \tag{1}$$

$$\vec{E} = \vec{E}_0 + \vec{E}_{ind}, \tag{2}$$

where $\vec{E}_0$ is *the applied electric field*, $\vec{u}$ is fluid velocity and $\vec{B}$ *total magnetic field*. In the generator, with no current flowing ($\vec{E}_0 = 0$), the equation is simplified as follows:

$$\vec{E} = \vec{E}_{ind} = \vec{u} \times \vec{B}. \tag{3}$$

The analysis of voltage, current, and power as functions of fluid velocity and magnetic field parameters is conducted using Faraday's law. Initially, by substituting Equation 4.3, the result is given by:

$$V_{oc} = -\int \vec{E}.dl = \int (\vec{u} \times \vec{B})\, dl = uBl, \tag{4}$$

where $l$ represents the characteristic distance. The *open circuit voltage* $V_{oc}$ is directly proportional to fluid velocity (u), magnetic field ($B$), and characteristic distance ($l$). *The current density* $\vec{J}$ can be calculated using:

$$\vec{J} = \sigma \vec{E} = \sigma uB, \tag{1}$$

and subtituting the equation to the current given by:

$$I = \int \vec{J}.dA = \sigma uBA, \tag{2}$$

Power ($P_e$) can be calculated by:

$$P_e = \int \vec{J}.\vec{E}\, dv = \sigma u^2 B^2 v, \tag{7}$$

where $v$ represent fluid volume. The electrical power is directly proportional to the electrical conductivity ($\sigma$), the square of the fluid velocity ($u^2$), the square of the magnetic field strength ($B^2$) and the volume ($v$). This relationship is derived from the open-circuit voltage ($V_{oc}$) where the load factor K=1 also referred to as the open circuit condition. The load factor ($K$) can be determined using Equation (8) and internal resistance ($R_i$) can be estimated using Equation (9).

$$K = \frac{R_{load}}{R_i + R_{load}} \tag{3}$$

$$R_i = \frac{l}{\sigma A}, \tag{4}$$

$R_{load}$ is external resistance and A is electrode total area. The load factor is defined as the ratio between the external resistance and the total resistance (the sum of external and internal resistance). The internal resistance ($R_i$) is directly proportional to the electrode spacing (l) and inversely proportional to the electrical conductivity ($\sigma$) and the total electrode area (A). This relationship can be derived through Ohm's Law, where the electric current (I) is first obtained using Equation (10),

$$I = \frac{V_{oc}}{R_i + R_{load}} \tag{5}$$

The relationship between the load factor (K) and the load voltage ($V_{load}$) can be substituted into Equation (8) and derived as

$$V_{oc} = I.(R_i + R_{load}) = I.\frac{R_{load}}{K} \tag{6}$$

$$V_{load} = I.R_{load} = K.V_{oc}, \tag{7}$$

Thus, the electrical power ($P_e$) is obtained using the following equation

$$P_e = V_{load}.I = K.V_{oc}.I \tag{8}$$

Ohm's Law involves resistance factors, which are expressed through the load factor (K) from Equation (10) to Equation (13). By combining Faraday's law with Ohm's law, the relationship between current (I) and electrical power ($P_e$) can be expressed as a function of fluid velocity (u), magnetic field strength (B), electrical conductivity (σ), and load factor (K), as well as the geometric variables of electrode spacing (l), electrode area (A), and chamber volume (v) of the MHD generator. By substituting Equation (4) and Equation (8) into Equation (11), the current (I) can be derived as follows:

$$I = \frac{uBl}{R_i+R_{load}} = V_{oc}\frac{1-K}{R_i} = uBl\frac{1-K}{R_i} \tag{9}$$

By substituting the internal resistance from Equation (9) into Equation (14), the following expression is obtained:

$$I = uBl\frac{1-K}{l/\sigma A} = \sigma uBA(1-K) \tag{10}$$

For the electrical power ($P_e$), by substituting Equation (5) and Equation (16) into Equation (13), the following expression is obtained:

$$P_e = K(uBl)(\sigma uBA(1-K)) = K(1-K)\sigma u^2 B^2 v, \tag{11}$$

By comparing the current (I) equations in Equation (7) and Equation (15), the multiplying factor (1−K) represents the resistance factor that governs the current. Similarly, in Equation (7) and Equation (13), the resistance factor for the electrical power ($P_e$) is expressed as K(1−K).

### III.2 Hydrodynamics Analysis

The two main equations describing fluid behaviour are the continuity equation and the Navier-Stokes equation. In cylindrical coordinates, these equations are detailed in Equations (17) to (20).

*Continuity Equation in cylindrical coordinates:*

$$\frac{1}{r}\frac{\partial}{\partial r}(r\,u_r) + \frac{1}{r}\frac{\partial}{\partial \theta}(u_\theta) + \frac{\partial}{\partial z}(u_z) = 0 \tag{12}$$

*Navier-Stokes Equation in cylindrical coordinates:*

Component r : $\rho\left[\frac{Du_r}{Dt} - \frac{u_\theta^2}{r}\right] = -\frac{\partial p}{\partial r} + \mu\left[\nabla^2 u_r - \frac{u_\theta}{r^2} - \frac{2}{r^2}\frac{\partial u_\theta}{\partial \theta}\right] + f_r$ (13)

Component θ : $\rho\left[\frac{Du_\theta}{Dt} + \frac{u_\theta u_r}{r}\right] = -\frac{1}{r}\frac{\partial p}{\partial \theta} + \mu\left[\nabla^2 u_\theta - \frac{u_\theta}{r^2} + \frac{2}{r^2}\frac{\partial u_r}{\partial \theta}\right] + f_\theta$ (14)

Component z : $\rho\frac{Du_z}{Dt} = -\frac{\partial p}{\partial z} + \mu\nabla^2 u_z + f_z$ (15)

The equations are simplified by making several assumptions: *steady state* ($\frac{Du}{Dt} = 0$) flow and velocity is primarily governed by the tangential component ($u_r = u_z = 0$). Since the flow is driven by a constant force, it is assumed that there is no change in velocity over time. The fluid movement is predominantly circular, leading to a second assumption that only the tangential component is considered in the equations. Given that the channel geometry at the data collection point is fixed, it is assumed that the main tangential flow is fully developed. With these assumptions, Equations (18) to (20) are simplified to:

Component r : $\rho\frac{u_\theta^2}{r} = \mu\frac{2}{r^2}\frac{\partial u_\theta}{\partial \theta}$ (21)

Component θ : $\frac{1}{r}\frac{\partial p}{\partial \theta} = \mu\left[\nabla^2 u_\theta - \frac{u_\theta}{r^2}\right] + f_\theta$ (22)

Component z : 0 (23)

Whereas external force $f_\theta$ is the force generated by electromagnetic interactions, which will be discussed in the following section.

### III.3 Magnetohydrodynamics Analysis

Based on its orientation shown in Fig. 1, the fluid flow is assumed to be solely tangential ($u_\theta$) and magnetic field is considered uniform in the axial direction ($B_z$). This configuration produced *current density* in the radial direction ($J_r$) according to Flemming's Right Hand Rule. The *current density* ($J_r$) interacts with magnetic field ($B_z$) generating Lorentz Force as described by the equation

$\vec{F_L} = \vec{J} \times \vec{B}$ (24)

Equation (24) is subtituted with *current density* ($\vec{J}$) from Equation (5), Lorentz Force becomes

$\vec{F_L} = \vec{J} \times \vec{B} = -[J_r B_z]\theta = -\sigma u_\theta B_z^2$ (25)

Thus, the Lorentz force acts on the fluid in the direction opposite to the fluid movement ($F_{L\theta}$). This force influences the fluid in the tangential component, resulting in the following Navier-Stokes equation for that component:

$$0 = -\frac{1}{r}\frac{\partial p}{\partial \theta} + \mu\left[\nabla^2 u_\theta - \frac{u_\theta}{r^2}\right] - \sigma u_\theta B_z^2 \qquad (26)$$

The Lorentz Force that moves against the direction of fluid causes a reduction in the force acting on the fluid. This can affect the fluid's speed and acceleration.

## IV    Results and Discussion

### IV.1    Mesh Independence Study

A mesh independence study was conducted to identify the optimal mesh size that balances accuracy and computational efficiency. The study evaluated fluid velocity between the peripheral and center electrodes at the axial midpoint. Table 2 shows the deviations across mesh sizes from coarse to fine. The normal mesh, with a 3% deviation was selected because it achieved good convergence without the high computational cost of the fine mesh.

Table 2.  Mesh Independence Study

| Mesh | Velocity magnitude (m/s) | Deviation |
| --- | --- | --- |
| coarser | 1.6091 | |
| coarse | 1.9366 | 20% |
| normal | 1.9872 | 3% |
| fine | 2.0212 | 2% |

### IV.2    Magnetic Field

The magnetic field is visualized in 3D using a plot of magnetic flux density, where the field lines originate from the north pole and extend toward the south pole. In the MHD generator, the electromagnetic induction interaction occurs within the inner regions of the disc and ring magnets. The inner side of the disc magnet acts as the south pole, while the ring magnet serves as the north pole, as illustrated in Fig. 4.

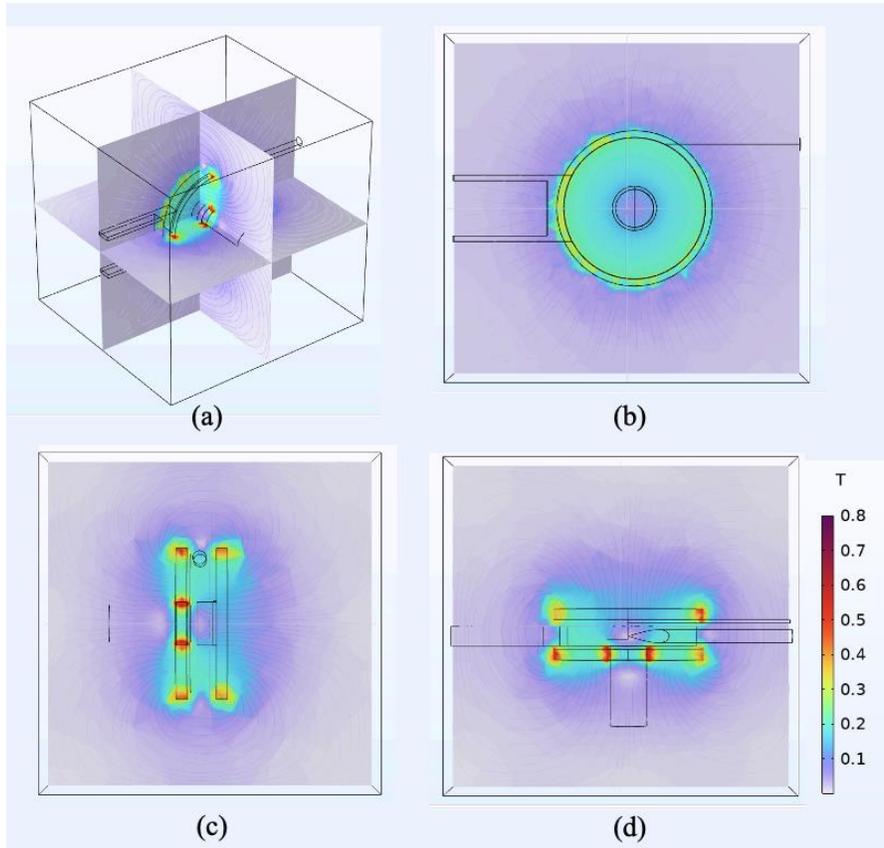

Figure 4. Magnetic Flux Density on (a) Isometric, (b) Front, (c) Left, and (d) Top view

## IV.3 Electric Potential

Fig. 5 illustrates the electric potential. The peripheral electrode serves as the ground, while the central electrode serves as the terminal. The difference in electric potential between these two electrodes represents the induced voltage.

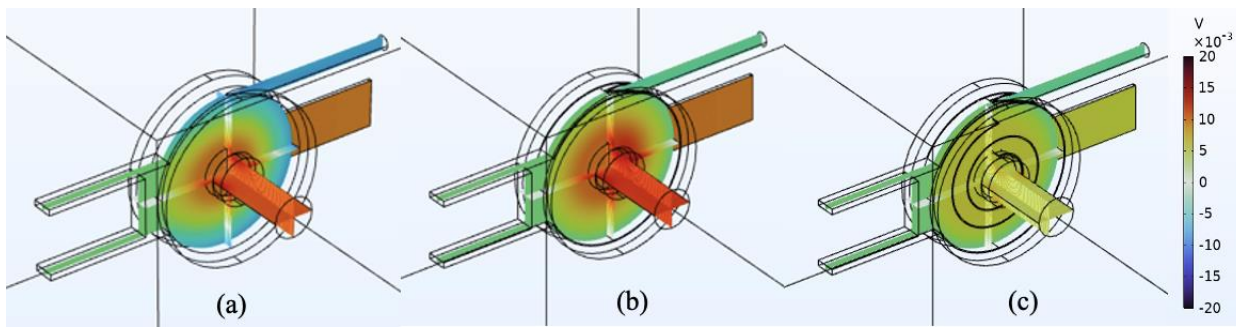

Figure 5. *Electric Potential* (V) on (a) Partial, (b) Whole, and (c) Spiral electrode

The peripheral electrode is grounded, meaning it has a negative charge and thus always has a potential of zero. In contrast, the central electrode has the highest potential and carries a positive charge.

## IV.4  Velocity, magnetic, electric, and current fields

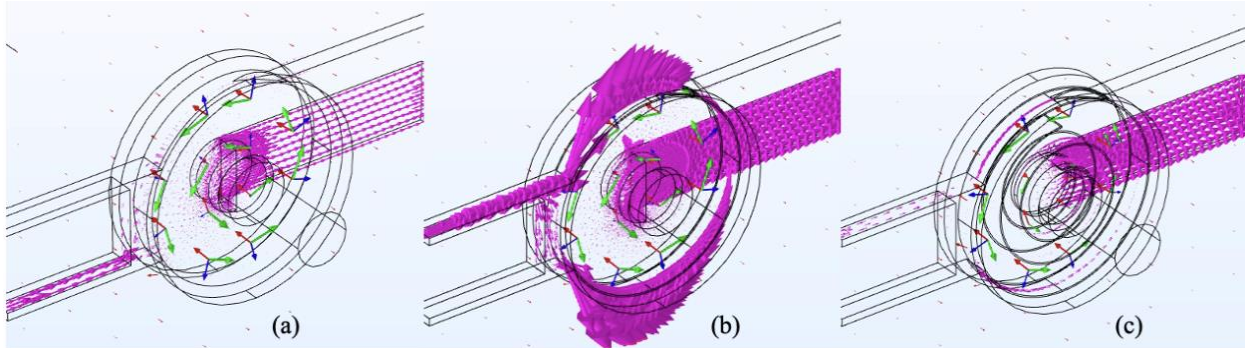

Figure 6. *Current Density* on (a) *Partial*, (b) *Whole* and (c) *Spiral electrode*

The fluid velocity, magnetic field, and electric field represented by green, red, and blue arrow, respectively. Fig. 6 illustrates the fields acting on the fluid. The magnetic field, represented in red, moves from the north (ring magnet) to the south (disc magnet). The fluid, shown as green arrows, flows tangentially, entering through the inlet and spiraling around the axial axis. The electric field, depicted by blue arrows, flows from the positive central electrode to the negative peripheral electrode. In the system under load, a current density flows from the positive to the negative electrode, represented by magenta arrows.

An electric potential present on the whole-area electrode surrounding the chamber. This occurs because the electrode serves as the surface where current flows. The same phenomenon is observed on the inner surface of the electrode, which is part of the central electrode in the design of the spiral electrode. The current flows from the positive electrode's outer side to the circuit and then to the negative electrode. With the increased surface area in the whole-area electrode configuration, more regions experience electromagnetic induction, resulting in a higher current output. The spiral wall within the chamber introduces additional friction losses due to the fluid's contact with the electrode surfaces. This leads to a significant reduction in velocity within the spiral region, which in turn affects both the voltage and power output. Although the spiral design increases the surface area and reduces internal resistance, the trade-off lies in the reduced distance between electrodes, which can lower the generated voltage.

## IV.5  Fluid Velocity

The velocity magnitude is visualized in a front view, whereas that the flow is predominantly tangential within the cylindrical coordinate system, as shown in Fig. 7.

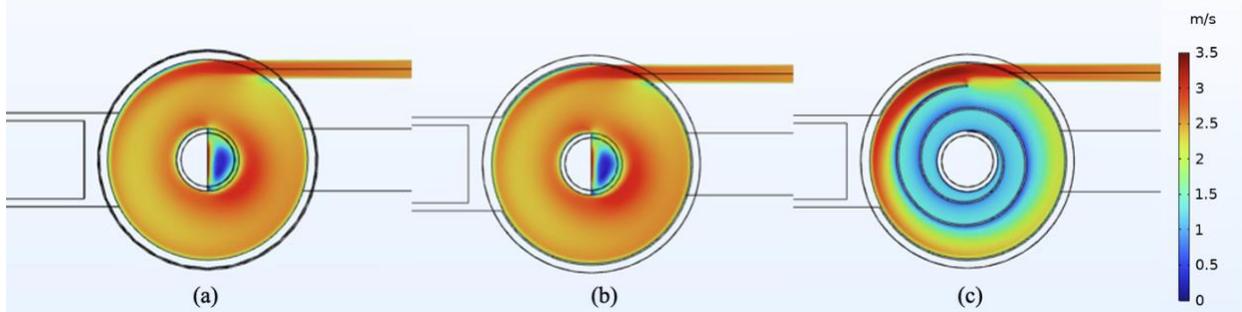

Figure 7. *Velocity Magnitude* on (a) Partial, (b) Whole, and (c) Spiral electrode

The highest velocity is observed when the flow exits the inlet and enters the chamber, where the cross-sectional area shifts from narrow to wide. Conversely, as the flow exits axially through the center outlet, the velocity approaches zero. This velocity change is attributed to the pressure drop, as explained by Equation (22). Boundary layers are observed on both the outer and inner walls because of the no-slip condition, resulting in near-zero velocities at those locations.

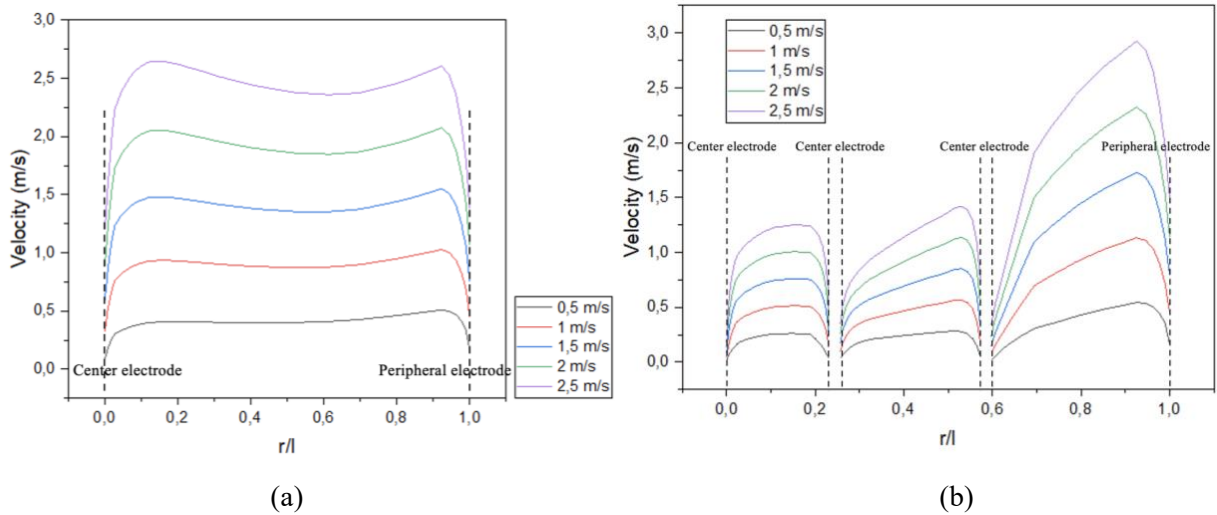

Figure 8. Velocity Profile (a) *Partial/Whole* and (b) *Spiral Electrode*

The velocity profiles exhibit a similar curve, where the lowest speed occurs near the walls or electrodes due to the no-slip condition caused by interaction with the solid surface. The velocity profile resembles an "M" shape, with a dip in the middle resulting from the Lorentz Force, which opposes the fluid flow. This velocity variation occurs along the $u_\theta$ component, as described in Equation (21). In the spiral electrode design, fluid velocity is reduced from the inlet as it interacts with the peripheral electrode at each turn, due to frictional forces with the central electrode's walls.

## IV.6 Electrical Performance

The open circuit electromotive force ($V_{oc}$) under no load of partial electrode configuration is plotted in Fig. Fig. 9. $V_{oc}$ against inlet velocity ($u_{inlet}$) for various magnetic field strength. These simulation values are compared to the theoretical $V_{oc}$ ($V_T$), which is calculated using Equation (4).

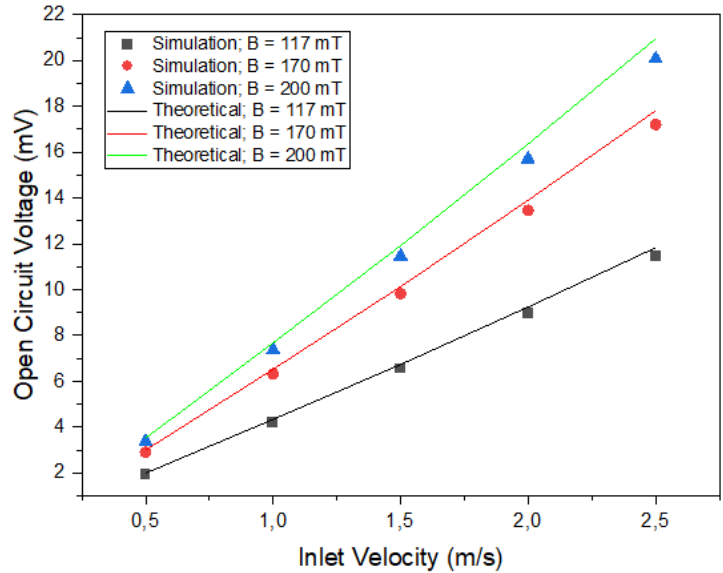

Figure 9. Comparison of open circuit voltage

The larger the inlet velocity, the higher the generated OCV. The difference between the theoretical and simulated OCV is relatively small, below 4% across all inlet velocity.

Table 3. Internal Resistance of Electrodes

| Geometry | Open circuit voltage (mV) | Loaded voltage (mV) | Current (mA) | External load (Ω) | Voltage drop (mV) | Internal resistance (Ω) |
|---|---|---|---|---|---|---|
| Partial | 4.214 | 1.545 | 0.399 | 3.87 | 2.669 | 6.69 |
| Whole | 4.243 | 1.499 | 1.034 | 1.45 | 2.744 | 2.65 |
| Spiral | 2.109 | 0.889 | 0.808 | 1.1 | 1.219 | 1.51 |

The whole-area electrode demonstrates middle ground of internal resistance due to its larger total area, which enhances its efficiency and power output. However, the spiral electrode, d, benefits from both a higher total area and an closer electrode spacing. This combination of factors allows the spiral electrode to achieve the lowest internal resistance, though its open circuit voltage is lower than that of the whole-area electrode.

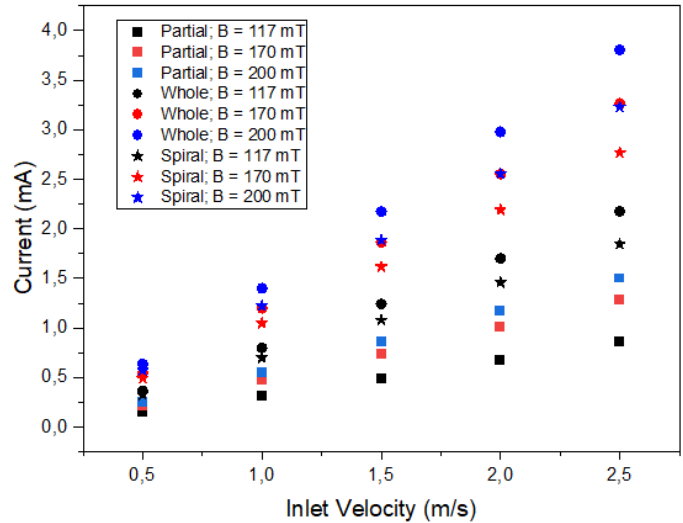

Figure 10. *Current* vs Inlet Velocity between Electrodes Geometry

Generally, the highest current is produced by the whole-area electrode, followed by the spiral electrode, and then the partial electrode. The current increases with a larger electrode area, making the whole-area electrode the most effective. Additionally, the current is directly proportional to the inlet velocity and the strength of the magnetic field.

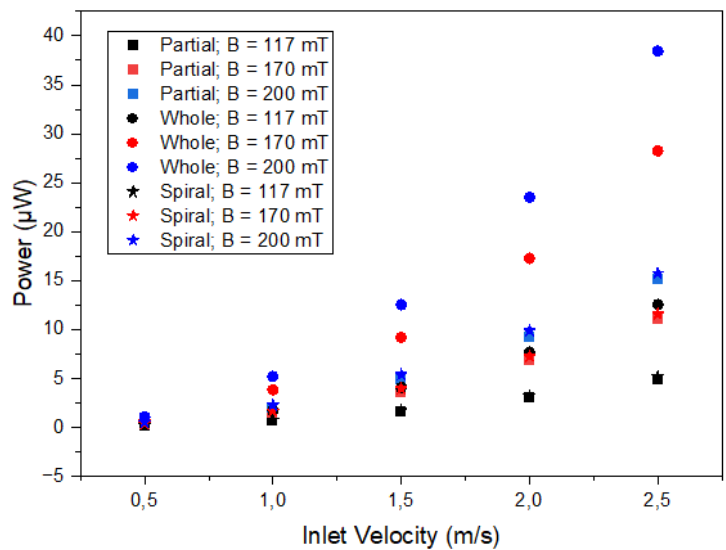

Figure 11. Power vs Inlet Velocity between Electrodes Geometry

Both electrode geometry enhancements show performance improvements, but the whole-area electrode delivers better results. By expanding the electrode area alone, the power output increased by 155% for the whole-area electrode. In contrast, the spiral electrode, which combines an increased surface area with closer electrode spacing, did not achieve a significant performance boost compared with the whole-area electrode.

Table 4. Electrical Performance Comparison between Electrodes

| Electrode | Partial (initial) | Whole | Spiral |
|---|---|---|---|
| $V_{load}$ (mV) | 10.039 | 10.092 | 4.8759 |
| I (mA) | 1.5005 | 3.8085 | 3.2291 |
| P (µW) | 15.064 | 38.435 | 15.745 |
| %increase (vs. partial geometry) |  | 155% | 5% |

## V   Conclusion

This study explored the performance of magnetohydrodynamics (MHD) generators with a focus on optimizing electrode geometry on vortex chamber design using seawater as the working fluid. Our investigation reveals that:

1. Electrode Geometry: Among the three configurations studied (partial, whole-area, and spiral), the whole-area electrode delivered the best performance due to its larger surface area, producing the highest open-circuit voltage ($V_{oc}$) and current. While offering advantages in electrode spacing and reduced internal resistance, the spiral electrode did not surpass the whole-area design in voltage output.

2. Internal resistance: The electrode area and spacing strongly influenced internal resistance. The whole-area electrode exhibited the lowest resistance, while the spiral design reduced resistance relative to the partial electrode but remained less effective overall.

3. Performance Metrics: The open-circuit voltage ($V_{oc}$) and current density were significantly affected by both inlet velocity and electrode design. The whole-area electrode consistently achieved the highest outputs, whereas the spiral electrode provided a more balanced trade-off between resistance reduction and current generation.

In summary, this study demonstrates that maximizing electrode area is critical to improving the performance of the MHD generator, with the whole-area design emerging as the most effective. The spiral configuration shows potential in reducing resistance, but its overall performance remains lower. Future work should explore advanced electrode geometries and fluid property optimization to further enhance the efficiency and practical application of MHD generators.

## Conflict of Interest

None declared.

**Acknowledgment:**

The authors acknowledge the financial support from Universitas Indonesia through Research Grant PUTI 2022-2023 (grant number: NKB-1487/UN2.RST/HKP.05.00/2022) is truly acknowledged.

## Data Availability

The data supporting the findings of this study are available from the corresponding author upon reasonable request.